\documentclass[sigconf,screen]{acmart}
\usepackage{graphicx}
\usepackage{bbding}
\usepackage{dsfont}
\usepackage{tablefootnote}
\usepackage{multirow,multicol}

\usepackage[dvipsnames]{xcolor}

\AtBeginDocument{%
  }

\copyrightyear{2026}
\acmYear{2026}
\setcopyright{cc}
\setcctype{by}
\acmConference[SIGIR '26] {Proceedings of the 49th International ACM SIGIR Conference on Research and Development in Information Retrieval}{July 20--24, 2026}{Melbourne, VIC, Australia.}
\acmBooktitle{Proceedings of the 49th International ACM SIGIR Conference on Research and Development in Information Retrieval (SIGIR '26), July 20--24, 2026, Melbourne, VIC, Australia}
\acmISBN{979-8-4007-2599-9/2026/07}
\acmDOI{10.1145/3805712.3808574}

\begin{document}

\title[A Reproducibility Study and Benchmarking of Counterfactual Explanations for Recommender Systems]{From Top-1 to Top-K: A Reproducibility Study and Benchmarking of Counterfactual Explanations for Recommender Systems}

\author{Quang-Huy Nguyen}
\authornote{Both authors contributed equally to this research.}
\orcid{0009-0006-7753-3843}
\affiliation{%
  \institution{VNU University of Engineering and Technology}
  \city{Hanoi}
  \country{Vietnam}
}
\email{22028077@vnu.edu.vn}

\author{Thanh-Hai Nguyen}
\authornotemark[1]
\orcid{0009-0009-0779-7028}
\affiliation{%
  \institution{VNU University of Engineering and Technology}
  \city{Hanoi}
  \country{Vietnam}
}
\email{23020057@vnu.edu.vn}

\author{Khac-Manh Thai}
\authornotemark[1]
\orcid{0009-0006-5213-2146}
\affiliation{%
  \institution{VNU University of Engineering and Technology}
  \city{Hanoi}
  \country{Vietnam}
}
\email{23021620@vnu.edu.vn}

\author{Duc-Hoang Pham}
\orcid{0009-0007-6847-384X}
\affiliation{%
  \institution{VNU University of Engineering and Technology}
  \city{Hanoi}
  \country{Vietnam}
}
\email{22021200@vnu.edu.vn}

\author{Huy-Son Nguyen}
\orcid{0009-0006-4616-0976}
\affiliation{%
 \institution{Delft University of Technology}
 \city{Delft}
 \country{The Netherlands}}
\email{H.S.Nguyen@tudelft.nl}

\author{Cam-Van Thi Nguyen}
\orcid{0009-0001-9675-2105}
\affiliation{%
  \institution{VNU University of Engineering and Technology}
  \city{Hanoi}
  \country{Vietnam}
}
\email{vanntc@vnu.edu.vn}

\author{Masoud Mansoury}
\orcid{0000-0002-9938-0212}
\affiliation{%
 \institution{Delft University of Technology}
 \city{Delft}
 \country{The Netherlands}}
\email{m.mansoury@tudelft.nl}

\author{Duc-Trong Le}
\orcid{0000-0003-4621-8956}
\affiliation{%
  \institution{VNU University of Engineering and Technology}
  \city{Hanoi}
  \country{Vietnam}
}
\email{trongld@vnu.edu.vn}

\author{Hoang-Quynh Le}
\orcid{0000-0002-1778-0600}
\affiliation{%
  \institution{VNU University of Engineering and Technology}
  \city{Hanoi}
  \country{Vietnam}
}
\email{lhquynh@vnu.edu.vn}

\renewcommand{\shortauthors}{Quang-Huy Nguyen et al.}

\begin{abstract}

Counterfactual explanations (CEs) provide an intuitive way to understand recommender systems by identifying minimal modifications to user-item interactions that alter recommendation outcomes. Existing CE methods for recommender systems, however, have been evaluated under heterogeneous protocols, using different datasets, recommenders, metrics, and even explanation formats, which hampers reproducibility and fair comparison. In this paper, we systematically reproduce, re-implement, and re-evaluate eleven state-of-the-art CE methods for recommender systems, covering both native explainers (e.g., LIME-RS, SHAP, PRINCE, ACCENT, LXR, GREASE) and specific graph-based explainers originally proposed for GNNs. Building on these implementations, a unified benchmarking framework is proposed to assess explainers along three dimensions: explanation format (implicit vs. explicit), evaluation level (item-level vs. list-level), and perturbation scope (user interaction vectors vs. user-item interaction graphs). Our evaluation protocol includes effectiveness, sparsity, and computational complexity metrics, and extends existing item-level assessments to top-K list-level explanations. Through extensive experiments on three real-world datasets and six representative recommender models, we analyze how well previously reported strengths of CE methods generalize across diverse setups. We observe that the trade-off between effectiveness and sparsity depends strongly on the specific method and evaluation setting, particularly under the explicit format; in addition, explainer performance remains largely consistent across item level and list level evaluations, and several graph-based explainers exhibit notable scalability limitations on large recommender graphs. Our results refine and challenge earlier conclusions about the robustness and practicality of CE generation methods in recommender systems, and provide a reproducible reference point for future work: \href{https://github.com/L2R-UET/CFExpRec}{https://github.com/L2R-UET/CFExpRec}.
\end{abstract}



\maketitle

\section{Introduction}

Recommender systems play a pivotal role in many large-scale applications due to their ability to identify items that align with users’ preferences and needs from an enormous pool of candidates \cite{zhang2019deep,hasan2024based}. Beyond recommendation accuracy, an increasingly important requirement is the provision of meaningful explanations for each recommendation, as explanations enhance user trust, system transparency, and interpretability \cite{peake2018explanation,zhang2020explainable}. Among the various forms of explanation, counterfactual explanations (CEs) have recently emerged as a particularly powerful paradigm \cite{saito2021counterfactual,tran2021counterfactual,barkan2024counterfactual,gurevitch2025lxr,chen2022grease,chen2025joint,ghazimatin2020prince,stratigi2026explaining}. By answering intuitive “\textit{what-if}” questions, such as how minimal changes in user behavior would alter a recommendation outcome, counterfactual explanations provide actionable and human-interpretable insights. From the user perspective, they improve understanding of the system’s decision logic, enable users to refine their interactions, and help obtain recommendations that more faithfully reflect their preferences. From the perspective of engineers and service providers, counterfactual explanations facilitate model debugging and diagnosis, and support the development of more robust and effective recommendation algorithms \cite{yao2022counterfactually,guidotti2024counterfactual}.

The need for reproducibility studies in the RecSys community is increasingly critical~\cite{shehzad2025we}. In pursuit of beyond-accuracy recommenders, while previous work has reproduced and examined issues related to fairness~\cite{nguyen2025reproducibility} and interpretability~\cite{dervishaj2025representation}, comprehensive studies on the evaluation of counterfactual explanations remain limited.
The current empirical evidence on CE generation methods in recommender systems is fragmented and difficult to compare in both their explanation paradigms and output formats. Based on the format of the explanation produced, existing approaches can be broadly categorized into two classes: \textit{implicit} methods, which imply the importance of each historical user-item interaction with respect to a target recommendation \cite{gurevitch2025lxr,barkan2024counterfactual}, and \textit{explicit} methods, which directly identify a subset of interactions whose removal would alter the recommendation outcome \cite{ghazimatin2020prince,tran2021counterfactual,chen2022grease,chen2025joint}. However, this heterogeneity in explanation targets has led to non-standardized evaluation protocols, as different methods adopt metrics tailored to their specific outputs. Consequently, performing fair and comprehensive comparisons of effectiveness across counterfactual explainers remains both necessary and challenging. Moreover, most existing methods are designed to explain a single recommended item, whereas modern recommender systems typically generate a ranked list of top-$K$ items. This discrepancy highlights the need for explanation and evaluation frameworks that also operate at the list level rather than focusing solely on individual recommendations. As a result, it is crucial to understand to what extent the reported strengths of individual CE generation methods are reproducible under a common protocol, how robust they are across datasets and recommender architectures, and whether their conclusions continue to hold when moving from top‑1 to full top‑$K$ recommendation lists.

In this paper, we take an initial yet significant step toward a comprehensive and systematic evaluation of CE generation methods for recommender systems by proposing a unified and reproducible benchmarking framework. Specifically, we formalize a standardized evaluation protocol that accommodates both implicit and explicit explanation formats, thereby ensuring fair, transparent, and consistent comparisons across diverse approaches. Under this framework, we collect a set of representative counterfactual explainers and benchmark them on multiple real-world datasets spanning different recommendation scenarios. To the best of our knowledge, this is the first study to systematically establish an evaluation protocol that operates at the level of the entire top-$K$ recommendation list. Moreover, we conduct an extensive empirical investigation to analyze the effectiveness, sparsity, robustness, and computational efficiency of different methods. Through this analysis, we uncover their relative strengths, inherent limitations, and practical trade-offs, providing actionable insights into how these explainers behave under various recommendation architectures and experimental conditions.

Owing to the inherently graph-structured nature of user-item interactions, graph neural network (GNN)-based recommender systems have been extensively studied and have consistently demonstrated superior performance compared to models that rely solely on direct user-item interactions in various domains~\cite{anand2025survey,nguyen2024bundle,nguyen2025ramen}. Despite these advances, most existing methods for generating CEs in recommendation focus on items with which users have directly interacted in their historical data, limiting their ability to produce faithful and informative explanations for recommender models that operate on the full user-item interaction graph. To bridge this gap and to provide a comprehensive benchmark for GNN-based recommenders, we adapt a wide range of GNN explainers~\cite{lucic2022cf,tan2022learning,ma2022clear,kang2024unr,ma2025c2explainer} originally developed for node-level and graph-level classification tasks to the recommendation setting. This adaptation facilitates a systematic comparison between recommender-specific explainers and general GNN-based explanation methods, and offers insights into the extent to which graph-aware explainers can capture the decision-making mechanisms of GNN-based recommenders.

In summary, the main contributions of our work are threefold:

\begin{itemize}
    \item We propose a unified and model-agnostic evaluation protocol for CEs in recommender systems, encompassing multiple explanation facets and enabling fair comparisons across methods. Notably, our framework is the first to evaluate CE effectiveness at the list level, moving beyond the conventional item-level assessment.
    
    \item We conduct extensive and reproducible experiments on three real-world datasets, benchmarking eleven CE generation methods across six representative recommender models. In particular, for graph-based recommenders, we adapt five GNN explainers originally designed for node and graph classification tasks to the recommendation setting, enabling a holistic and systematic comparison.
    
    \item We provide a comprehensive empirical analysis from multiple perspectives, offering insights into the effectiveness, robustness, and computational complexity of different counterfactual explanation approaches under varying recommender architectures and data characteristics.
\end{itemize}

\section{Preliminaries and Notation}

Let $\mathcal{U}$ and $\mathcal{I}$ denote the sets of users and items, respectively, and let $\mathcal{E}$ denote the set of user-item interactions. We model the system as a bipartite graph $\mathcal{G} = (\mathcal{U} \cup \mathcal{I}, \mathcal{E})$, which can be represented by a binary interaction matrix $\mathbf{X} \in \{0,1\}^{|\mathcal{U}| \times |\mathcal{I}|}$. 
Each row $\mathbf{x}_u \in \{0,1\}^{|\mathcal{I}|}$ corresponds to the interaction history of user $u$, where $\mathbf{x}_u[i] = 1$ indicates that user $u$ has previously interacted with item $i$. 
We denote by $\mathcal{I}_u$ the set of items with which user $u$ has interacted. Throughout this paper, the term \textit{interaction} is used interchangeably to refer to an item in $\mathcal{I}_u$ or, equivalently, an edge in the graph $\mathcal{G}$.

We consider a trained recommender model $f_\theta$ with parameters $\theta$. For simplicity, throughout the paper, we focus on the mathematical formulation with regard to models that take an individual user interaction vector $\mathbf{x}_u$ as input, noting that the formulation naturally extends to models operating in the full matrix $\mathbf{X}$. Given $\mathbf{x}_u$, the model assigns a relevance score $f_\theta(i,\mathbf{x}_u)$ to each candidate item $i\in\mathcal{I}\setminus\mathcal{I}_u$. Items are ranked in descending order of these scores, and the ranking position of item $i$ is denoted by $\mathrm{rank}(i;\mathbf{x}_u)\in\{1,2,...,|\mathcal{I}_u|\}$. In the scope of this work, we restrict our attention to the top-$K$ recommendations, i.e., items that satisfy $\mathrm{rank}(i;\mathbf{x}_u)\le K$. Since our objective is to explain the model's behavior, the model parameters remain fixed throughout the explanation process.

CE generation methods assess the sensitivity of the top-$K$ recommendation list to input perturbations. In the \emph{implicit} format, the explainer produces an explanation mask $\mathbf{m}\in[0,1]^{|\mathcal{I}|}$, where each entry reflects the importance of the corresponding historical interaction. During evaluation, items in $\mathcal{I}_u$ are progressively removed by setting their corresponding entries in $\mathbf{x}_u$ to zero, following the ranking of their explanation scores. Removing items with the highest scores yields a sequence of perturbed inputs $\{\mathbf{x}_u^{(t,\text{pos})}\}$, whereas removing those with the lowest scores produces $\{\mathbf{x}_u^{(t,\text{neg})}\}$. In contrast, \emph{explicit} CEs directly add or remove interactions in the user profile, resulting in a perturbed vector $\mathbf{x}_u^{p}\in\{0,1\}^{|\mathcal{I}|}$. In both cases, explanation is measured by the extent to which these perturbations alter the resulting top-$K$ recommendations.

\section{Benchmarking Framework}
\label{sec:benchmark_framework}

\begin{table*}[!ht]
    \centering
    \caption{Comparison of evaluation protocol with different CE generation methods for recommender systems. The symbol \checkmark denotes the type of evaluation that the method uses in the original paper.}
    \resizebox{\textwidth}{!}{
    \begin{tabular}{lcccccccccccc}
        \hline
        \multirow{2}{*}{\textbf{Paper}} & \multicolumn{2}{c}{\textbf{Explanation Format}} & & \multicolumn{2}{c}{\textbf{Evaluation Level}} & & \multicolumn{2}{c}{\textbf{Perturbation Scope}} & & \multicolumn{3}{c}{\textbf{Evaluation Criteria}}\\
        \cline{2-3} \cline{5-6} \cline{8-9} \cline{11-13}
        & \textbf{Implicit} & \textbf{Explicit} & & \textbf{Item-level} & \textbf{List-level} & & \textbf{User Vector} & \textbf{Graph} & & \textbf{Effectiveness} & \textbf{Sparsity} & \textbf{Complexity}\\ 
        \hline
        LIME-RS \cite{nobrega2019towards} & \checkmark & & & \checkmark & & & \checkmark & & & \checkmark & & \\
        SHAP \cite{zhong2022shap} & \checkmark & & & \checkmark & \checkmark & & \checkmark & & & \checkmark & \checkmark & \checkmark \\
        PRINCE \cite{ghazimatin2020prince} & & \checkmark & & \checkmark & & & \checkmark & & & \checkmark & \checkmark & \checkmark \\
        ACCENT \cite{tran2021counterfactual} & \checkmark & \checkmark & & \checkmark & & & \checkmark & & & \checkmark & \checkmark & \\
        LXR \cite{barkan2024counterfactual,gurevitch2025lxr} & \checkmark & & & \checkmark & & & \checkmark & & & \checkmark & & \checkmark \\
        GREASE \cite{chen2022grease,chen2025joint} & & \checkmark & & \checkmark & & & & \checkmark & & \checkmark & \checkmark & \checkmark \\
        
        \hline
        \textbf{Ours} & $\boldsymbol{\checkmark}$ & $\boldsymbol{\checkmark}$ & &  $\boldsymbol{\checkmark}$ & $\boldsymbol{\checkmark}$ & & $\boldsymbol{\checkmark}$ & $\boldsymbol{\checkmark}$ & & $\boldsymbol{\checkmark}$ & $\boldsymbol{\checkmark}$ & $\boldsymbol{\checkmark}$\\
        \hline
    \end{tabular}
    }
    \label{tab:method_overview}
\end{table*}

This section outlines the design of our reproducibility and benchmarking framework. We first introduce the benchmarking protocol, defined along key dimensions that differentiate existing explainers, including explanation format, evaluation level, and perturbation scope. We then present the evaluation criteria and corresponding metrics for assessing effectiveness from multiple perspectives. Finally, we describe the implementation details, including datasets, compared explainers, underlying recommender models, and experimental settings. A comparison with prior CE evaluation protocols is provided in Table~\ref{tab:method_overview}.

\subsection{Benchmarking Protocol}

\paragraph{\textbf{Explanation Format}} Our work aims to evaluate different counterfactual explainers under two distinct formats of the produced explanations. In the \emph{implicit format}, the CE model produces importance scores that quantify the contribution of each interaction to the recommendation outcome. In the \textit{explicit format}, the CE model directly outputs a set of user-item interactions to perturb, with the goal of inducing a change in the recommended list.
\paragraph{\textbf{Evaluation Level}} Regarding changes in recommendation outcomes, we evaluate CE methods through a top-$K$ recommendation list using two evaluation levels: \textit{item-level evaluation} and\textit{ list-level evaluation}. In the\textit{ item-level evaluation}, the goal of the counterfactual explainer is to generate a CE that results in the removal of a specific target item from the top-$K$ recommendation list. Consequently, for a list of top-$K$ items, the model produces $K$ distinct counterfactual explanations, each corresponding to a different recommended item. Meanwhile, under the \textit{list-level evaluation}, the explainer seeks a single counterfactual intervention that causes all items in the top-$K$ list to be removed simultaneously. In this case, the model outputs a unified CE for each top-$K$ recommendations.
\paragraph{\textbf{Perturbation Scope}} Recommender models differ in their input representations, ranging from methods that rely solely on \textit{a user’s historical interaction vector} to more complex architectures that explicitly model the\textit{ full user-item interaction graph}. Correspondingly, counterfactual explainers also vary in the scope of interactions they are permitted to perturb. To ensure a comprehensive and fair evaluation, our benchmark includes both categories of recommender models and enables a holistic comparison of counterfactual explanation methods across these architectural paradigms.

\subsection{Benchmarking Criteria and Metrics}

In our study, we define a set of key criteria to evaluate the quality of counterfactual explainers, and use them to design evaluation protocols and metrics for empirical analysis. Specifically, we assess performance according to three different criteria: effectiveness, explanation sparsity, and computational complexity.

\paragraph{\textbf{Effectiveness}}

Effectiveness measures the extent to which CE can alter the ranking of a single item or modify the top-$K$ recommendation list. Under the implicit format, a CE is considered highly effective if it prioritizes interactions whose removal most effectively degrades the recommendation outcome. To quantify this property, we adopt two widely used perturbation-based metrics: \textbf{Positive Perturbation (POS-P)} and \textbf{Negative Perturbation (NEG-P)} \cite{barkan2024counterfactual,gurevitch2025lxr,mohammadi2025beyond}. Both metrics evaluate how quickly a target item drops out of the top-$K$ list as input interactions are progressively masked. Specifically, POS-P removes interactions in decreasing order of importance, whereas NEG-P removes them in increasing order of importance. A lower POS-P indicates that removing highly important interactions rapidly leads to decline in item ranking, while a higher NEG-P indicates that removing low-importance interactions has minimal effect. Together, these behaviors reflect stronger sufficiency of the explanation.
Detailed formulations of POS-P with regard to an explanation for a pair of user $u$ and item $i$ in top-$K$ recommendation (item-level evaluation) and for a user $u$ (list-level evaluation) are described as follows:
\begin{gather}
    \text{POS-P@K}(u-i) = \frac{1}{T}\sum_{t=1}^T\mathds{1}\Big[\text{rank}(i;\mathbf{x}_u^{(t,pos)})\leq K\Big],\\
        \text{POS-P@K}(u) = \frac{1}{TK}\sum_{t=1}^T\sum_{k=1}^K\mathds{1}\Big[\text{rank}(i_k;\mathbf{x}_u^{(t, pos)})\leq K\Big],
\end{gather}
where $T$ denotes the number of perturbation steps during evaluation, $i_k$ represents the item in position $k$ of recommended list before CE is applied, and $\mathds{1}[\cdot]$ represents the indicator function, which equals $1$ if the condition inside the brackets holds and $0$ otherwise. At each perturbation step, we remove $(100/T)\%$ of the interactions. NEG-P is defined analogously, with $\mathbf{x}_u^{(t,pos)}$ replaced by $\mathbf{x}_u^{(t,neg)}$.

Under the explicit format, CE is considered effective with respect to a single item if it successfully removes the item from the top-$K$ recommendation list. When explaining an entire top-$K$ list, explainer effectiveness further depends on how many items are displaced and how prominently the newly introduced items appear in the ranking (that is, replacing items at higher-ranked positions reflects a stronger impact on the recommendation outcome). 

Based on this formulation, we follow the evaluation conducted in ~\cite{chen2025joint} and some papers for general GNN explainers \cite{tan2022learning,kang2024unr} and introduce the \textbf{Probability of Necessity - Single (PN-S)}. For single-item explanations, this metric measures the proportion of cases in which the CE causes the target item to drop out of the top-$K$. For top-$K$ list explanations, PN-S is extended to compute the ratio of items removed from the list, capturing the overall degree of change. Mathematically, PN-S with regard to an explanation for a pair of user $u$ and item $i$ in top-$K$ recommendation (item-level evaluation) and for a user $u$ (list-level evaluation) are formulated as follows:
\begin{gather}
    \text{PN-S@K}(u-i) = \mathds{1}\Big[\text{rank}(i;\mathbf{x}_u^{p})>K\Big]\\
    \text{PN-S@K}(u) = \frac{1}{K}\sum_{k=1}^K\mathds{1}\Big[\text{rank}(i_k;\mathbf{x}_u^{p})>K\Big]
\end{gather}

To further account for ranking sensitivity, when evaluating for full top-$K$ item list, we introduce a new metric, \textbf{Probability of Necessity - Ranking (PN-R)}, inspired by the position-aware design of NDCG metric used in recommendation evaluation. Unlike PN-S, which treats all ranks equally, PN-R assigns higher weights to changes occurring at higher-ranked positions in the top-$K$. Concretely, when a new item enters the top-$K$, it receives a position-dependent score, with larger rewards for occupying higher ranks. This design reflects the intuition that a CE is more sufficient if it causes substantial changes in recommendation of higher position, thereby indicating a stronger and more meaningful change of the model’s output. PN-R is mathematically formulated with regard to a user $u$ in list-level evaluation as follows:
\begin{equation}
    \text{PN-R@K}(u) = 1-\dfrac{\displaystyle\sum\limits_{k=1}^K\dfrac{\mathds{1}\Big[\text{rank}(i_k;\mathbf{x}_u^p)\leq K\Big]}{\text{log}_2\Big(\text{rank}(i_k;\mathbf{x}_u^p)+1\Big)}}{\displaystyle\sum\limits_{k=1}^K\dfrac{1}{\text{log}_2(k+1)}}
\end{equation}

\paragraph{\textbf{Explanation Sparsity}}

Beyond sufficiency, explanation sparsity is a key criterion for evaluation quality. High-quality explanations should rely on a minimal set of interactions rather than dispersing importance across many weakly relevant ones. Greater sparsity enhances interpretability, reduces cognitive burden, and improves actionability by requiring fewer changes to alter predictions.

For the implicit format, we measure sparsity using the \textbf{Gini Index} \cite{hurley2009comparing}, which quantifies how concentrated the explanation is across interactions, with greater inequality (i.e., a higher score) indicating that the explanation concentrates on a smaller number of salient signals, thereby reflecting superior explanatory quality. This makes the Gini Index well-suited for capturing the degree of focus in attribution-based explanations. The formulation for Gini Index for an explanation is illustrated as follows:
\begin{equation}
    \text{Gini} = 1-2\sum_{k=1}^{|\mathcal{I}_u|}\left(\dfrac{\mathbf{\hat{m}}_s[k]}{\|\mathbf{\hat{m}}_s\|_1}\cdot\dfrac{|\mathcal{I}_u|-k+0.5}{|\mathcal{I}_u|}\right),
\end{equation}
where $\mathbf{\hat{m}}_s$ represents the importance scores of all interactions, normalized via min-max scaling and arranged in ascending order.

For the explicit format, we directly compute the number of perturbations (\textbf{\#Perturb}) applied to the original input, as this provides a straightforward and interpretable measure of how minimal the counterfactual intervention is. The mathematical formulation of this metric for an explanation is as follows:
\begin{equation}
    \text{\#Perturb} = \sum_{i\in\mathcal{I}_u}\mathds{1}\left[\mathbf{x}_u^p[i]\neq\mathbf{x}_u[i]\right]
\end{equation}

\paragraph{\textbf{Computational Complexity}}
Computational complexity is determined by the computational resources and time required to estimate an individual CE. This criterion is essential, as practical deployment in real-world applications necessitates that the proposed methods remain computationally feasible. Accordingly, our work reports the inference time required to compute a CE.

\subsection{Recommendation Models}
We consider three recommendation models that operate on user vector representations. \textbf{(1) MF} \cite{koren2009matrix} is a classical latent factor model that use \textit{matrix factorization} technique to learn low-dimensional embeddings for users and items. \textbf{(2) VAE} \cite{liang2018variational} employs \textit{variational autoencoder} to learn probabilistic latent representations through variational inference, enabling effective modeling of uncertainty and complex user preferences. \textbf{(3) DiffRec} \cite{wang2023diffusion} formulates recommendation as a \textit{denoising diffusion process} over user representations, progressively refining noisy latent variables to generate high-quality recommendations.
 
We further include three graph-based recommender systems that explicitly leverage the structure of the user-item interaction graph. \textbf{(4) LightGCN} \cite{he2020lightgcn} is a simplified version of \textit{graph convolutional network}, which removes nonlinear transformations and feature transformations, focuses solely on neighborhood aggregation to learn collaborative filtering signals. \textbf{(5) GFormer} \cite{li2023graph} incorporates \textit{transformer}-based architectures into graph recommendation, enabling long-range dependency modeling and adaptive weighting of neighborhood information. \textbf{(6) SimGCL} \cite{yu2022graph} adopts a \textit{contrastive learning} framework that introduces stochastic graph augmentations to improve representation robustness and generalization in graph-based collaborative filtering.

\subsection{Counterfactual Explainers}

\subsubsection{Native Counterfactual Explainers for Recommendation Model}

For native counterfactual explainers tailored to recommender models, we consider six representative methods for comparison. \textbf{(1) LIME-RS} \cite{nobrega2019towards} is an adaptation of the model-agnostic, post hoc explanation method LIME \cite{ribeiro2016should}, which approximates a model’s local decision boundary using a simple, interpretable linear surrogate model. \textbf{(2) SHAP} \cite{zhong2022shap} quantifies individual feature contributions to model predictions using Shapley values from cooperative game theory, reflecting each feature’s marginal impact on the final recommendation. \textbf{(3) PRINCE} \cite{ghazimatin2020prince} proposes a polynomial-time algorithm based on Personalized PageRank \cite{haveliwala2002topic} and random walk to identify a minimal set of counterfactual actions. \textbf{(4) ACCENT} \cite{tran2021counterfactual} is a counterfactual explanation framework that extends influence function-based techniques \cite{koh2017understanding,cheng2019incorporating} to neural recommender systems, providing actionable and model-agnostic explanations. \textbf{(5) LXR} \cite{barkan2024counterfactual,gurevitch2025lxr} employs a self-supervised learning paradigm with mask optimization to generate explanation masks that identify critical components of user data. To generate CE with explicit format, we apply a thresholding strategy, in which items with an importance score greater than 0.5 are identified as CE for the user. \textbf{(6) GREASE} \cite{chen2022grease} is a counterfactual explainer for GNN-based recommenders that trains a surrogate model and optimizes structured interventions on the user-item graph to identify counterfactual factors.
\subsubsection{GNN-based Counterfactual Explainers}
To enable a comprehensive comparison with GNN-based recommendation models, we further adapt five counterfactual explanation methods originally proposed for GNNs in node and graph classification tasks to generate CEs within the recommendation setting. \textbf{(7) CF-GNNExplainer} \cite{lucic2022cf} is a counterfactual framework for node classification that identifies a critical subgraph and applies gradient-based optimization to find minimal perturbations that flip predictions. We adapt it by redefining the flipping condition as item removal from the top-$K$ (item-level) or a change in the top-$K$ list (list-level). \textbf{(8) CF$^2$} \cite{tan2022learning} generates explanations by jointly modeling factual and counterfactual reasoning through a composite loss function. In our setting, we isolate and utilize only the counterfactual component of the objective. \textbf{(9) CLEAR} \cite{ma2022clear} employs a graph variational autoencoder to generate CEs. In addition, CLEAR introduces auxiliary variables to better model the underlying causal structure of graph data. In our setting, we follow the node classification paradigm suggested in the original paper, where each interaction is associated with an auxiliary variable to capture the causal structure of the user-item graph. \textbf{(10) UNR-Explainer} \cite{kang2024unr} is designed for unsupervised node representation learning and generates CEs by estimating subgraph importance via Monte Carlo Tree Search. In our setting, we define the subgraph importance score as the proportion of items removed from the top-$K$ list after removing the subgraph. \textbf{(11) C2Explainer} \cite{ma2025c2explainer} adopts a mask-based explanation framework and employs a straight-through estimator to address the issue of “introduced evidence” caused by fractional masks. It further supports edge addition during counterfactual graph construction. In our experiments, we evaluate both the original version and a variant without edge addition ("-add" variant).

\subsection{Datasets}

We evaluate the proposed framework on three datasets with varying scales and sparsity levels. Following \cite{barkan2024counterfactual,gurevitch2025lxr}, we use ML1M \cite{10.1145/2827872} and Yahoo \cite{10.5555/3000375.3000377} for explainer evaluation, and additionally include the Amazon Fashion dataset from the Amazon Reviews 2023 collection \cite{hou2024bridging}, hereafter referred to as Amazon. For Amazon, we construct an implicit feedback setting by retaining ratings greater than 3 as positive interactions and removing the rest, followed by filtering out users and items with fewer than three interactions to mitigate extreme sparsity. Dataset statistics are summarized in Table \ref{tab:dataset}.

\begin{table}[t!]
    \setlength{\tabcolsep}{3.7pt}
    \centering
    \caption{Summary of Dataset Statistics}
    \begin{tabular}{lcccccc}
        \hline
        \textbf{Dataset} & \textbf{$|\mathcal{U}|$} & \textbf{$|\mathcal{I}|$} & 
        \textbf{$|\mathcal{E}|$} & \textbf{$|\mathcal{E}|/|\mathcal{I}|$} & \textbf{$|\mathcal{E}|/|\mathcal{U}|$} & \textbf{Sparsity} \\
        \hline
        Amazon & 1,275 & 1,374 & 6,973 & 5.07 & 5.46 & 99.6020\% \\
        ML1M & 6,037 & 3,381 & 575,128 & 170.16 & 95.27 & 97.1823\% \\
        Yahoo & 13,797 & 4,604 & 365,750 & 26.51 & 78.83 & 99.4242\% \\
        \hline
    \end{tabular}
    \label{tab:dataset}
\end{table}

\subsection{Experimental Details}
For recommender training, each dataset is split into training, validation, and test sets with an 8:1:1 ratio. Early stopping is applied if validation performance does not improve for 20 consecutive epochs, ensuring selection of the best-performing model. For explanation generation, we randomly sample 500 users for evaluation, while the remaining users are used for training in methods requiring pretraining (i.e., LXR and CLEAR). During evaluation, we consider top-3 and top-5 recommendation lists. Under the implicit format, the number of perturbation steps $T$ is fixed at 10 across all experiments. The final performance is reported as the average score together with standard deviation across all generated explanations.

All experiments are implemented in PyTorch\footnote{\href{https://docs.pytorch.org/docs/stable/index.html}{https://docs.pytorch.org/docs/stable/index.html}} and conducted on two NVIDIA T4 GPUs. For both recommendation and explanation models, we either reconstruct the methods or reuse publicly available implementations when possible. Given that our experimental setup differs from those adopted in the original studies, we re-tune the hyperparameters using Optuna\footnote{\href{https://optuna.readthedocs.io/en/stable/}{https://optuna.readthedocs.io/en/stable/}} framework to identify the optimal configurations with respect to efficiency metrics (i.e. POS-P and PN-S), and re-evaluate their performance under our proposed benchmarking framework to ensure consistency and fairness in comparison. To facilitate reproducibility, we release all implementations and experimental configurations in a public repository\footnote{\href{https://github.com/L2R-UET/CFExpRec}{https://github.com/L2R-UET/CFExpRec}}.

\section{Empirical Investigation}

\begin{table*}[!t]
    \centering
    \setlength{\tabcolsep}{4pt}
    \caption{Performance comparison under the implicit format with MF as recommendation model (user interaction vector as input). P@K represents POS-P@K, N@K represents NEG-P@K, G@K represents Gini@K. \textbf{Bold} represents the best explainer. \underline{Underline} represents the second best explainer. $\uparrow$ indicates that higher values correspond to better performance, whereas $\downarrow$ indicates that lower values correspond to better performance.}
    \resizebox{\textwidth}{!}{
    \begin{tabular}{llccccccccccccc}
    \hline
    \multirow{2}{*}{\textbf{Dataset}} & \multirow{2}{*}{\textbf{Model}} & \multicolumn{6}{c}{\textbf{Item-level}} & & \multicolumn{6}{c}{\textbf{List-level}} \\\cline{3-8} \cline{10-15}
    & & \textbf{P@3$\downarrow$} & \textbf{N@3$\uparrow$} & \textbf{G@3$\uparrow$} & \textbf{P@5$\downarrow$} & \textbf{N@5$\uparrow$} & \textbf{G@5$\uparrow$} & & \textbf{P@3$\downarrow$} & \textbf{N@3$\uparrow$} & \textbf{G@3$\uparrow$} & \textbf{P@5$\downarrow$} & \textbf{N@5$\uparrow$} & \textbf{G@5$\uparrow$} \\
    \hline
    \multirow{8}{*}{Amazon} & \multirow{2}{*}{LIME-RS} & 0.1471 & 0.1495 & 0.4256 & 0.1682 & 0.1719 & \underline{0.4258} & & 0.1449 & 0.1485 & 0.4149 & 0.1674 & 0.1711 & 0.4157 \\[-0.7ex]
    & & \footnotesize{($\pm$0.2260)} & \footnotesize{($\pm$0.2337)} & \footnotesize{($\pm$0.1393)} & \footnotesize{($\pm$0.2497)} & \footnotesize{($\pm$0.2553)} & \footnotesize{($\pm$0.1399)} & & \footnotesize{($\pm$0.1432)} & \footnotesize{($\pm$0.1422)} & \footnotesize{($\pm$0.1361)} & \footnotesize{($\pm$0.1354)} & \footnotesize{($\pm$0.1342)} & \footnotesize{($\pm$0.1369)} \\
    & \multirow{2}{*}{SHAP} & \underline{0.1312} & \underline{0.1575} & 0.1396 & \underline{0.1570} & \underline{0.1798} & 0.1525 & & \underline{0.1387} & \underline{0.1538} & 0.1685 & \underline{0.1637} & \underline{0.1726} & 0.1878 \\[-0.7ex]
    & & \footnotesize{($\pm$0.2204)} & \footnotesize{($\pm$0.2328)} & \footnotesize{($\pm$0.2356)} & \footnotesize{($\pm$0.1421)} & \footnotesize{($\pm$0.1435)} & \footnotesize{($\pm$0.2356)} & & \footnotesize{($\pm$0.1421)} & \footnotesize{($\pm$0.2532)} & \footnotesize{($\pm$0.2692)} & \footnotesize{($\pm$0.1287)} & \footnotesize{($\pm$0.1364)} & \footnotesize{($\pm$0.2823)} \\
    & \multirow{2}{*}{ACCENT} & 0.1491 & 0.1453 & \textbf{0.4818} & 0.1696 & 0.1688 & 0.1685 & & 0.1570 & 0.1429 & \underline{0.4784} & 0.1733 & 0.1660 & \underline{0.4746} \\[-0.7ex]
    & & \footnotesize{($\pm$0.2273)} & \footnotesize{($\pm$0.2315)} & \footnotesize{($\pm$0.2248)} & \footnotesize{($\pm$0.2493)} & \footnotesize{($\pm$0.2546)} & \footnotesize{($\pm$0.2606)} & & \footnotesize{($\pm$0.2488)} & \footnotesize{($\pm$0.1391)} & \footnotesize{($\pm$0.2244)} & \footnotesize{($\pm$0.1349)} & \footnotesize{($\pm$0.1328)} & \footnotesize{($\pm$0.2232)} \\
    & \multirow{2}{*}{LXR} & \textbf{0.1161} & \textbf{0.1805} & \underline{0.4809} & \textbf{0.1481} & \textbf{0.1916} & \textbf{0.4769} & & \textbf{0.1189} & \textbf{0.1806} & \textbf{0.4795} & \textbf{0.1482} & \textbf{0.1895} & \textbf{0.4802} \\[-0.7ex]
    & & \footnotesize{($\pm$0.2154)} & \footnotesize{($\pm$0.2475)} & \footnotesize{($\pm$0.1733)} & \footnotesize{($\pm$0.2417)} & \footnotesize{($\pm$0.2647)} & \footnotesize{($\pm$0.1688)} & & \footnotesize{($\pm$0.1288)} & \footnotesize{($\pm$0.1613)} & \footnotesize{($\pm$0.1704)} & \footnotesize{($\pm$0.1211)} & \footnotesize{($\pm$0.1478)} & \footnotesize{($\pm$0.1643)} \\
    \hline
    \multirow{8}{*}{ML1M} & \multirow{2}{*}{LIME-RS} & \underline{0.1513} & \textbf{0.5995} & 0.2700 & \underline{0.1660} & \textbf{0.6348} & 0.2699 & & \underline{0.2355} & \textbf{0.4549} & 0.2695 & \underline{0.2738} & \textbf{0.4569} & 0.2683 \\[-0.7ex]
    & & \footnotesize{($\pm$0.2503)} & \footnotesize{($\pm$0.2697)} & \footnotesize{($\pm$0.0658)} & \footnotesize{($\pm$0.2651)} & \footnotesize{($\pm$0.2629)} & \footnotesize{($\pm$0.0652)} & & \footnotesize{($\pm$0.1828)} & \footnotesize{($\pm$0.1639)} & \footnotesize{($\pm$0.0700)} & \footnotesize{($\pm$0.1611)} & \footnotesize{($\pm$0.1402)} & \footnotesize{($\pm$0.0631)} \\
    & \multirow{2}{*}{SHAP} & 0.1807 & 0.4978 & \underline{0.3948} & 0.2181 & 0.4956 & \underline{0.4799} & & 0.2687 & 0.3788 & \underline{0.5166} & 0.2800 & 0.4153 & \underline{0.4240} \\[-0.7ex]
    & & \footnotesize{($\pm$0.2502)} & \footnotesize{($\pm$0.2865)} & \footnotesize{($\pm$0.1317)} & \footnotesize{($\pm$0.2801)} & \footnotesize{($\pm$0.3113)} & \footnotesize{($\pm$0.1217)} & & \footnotesize{($\pm$0.1875)} & \footnotesize{($\pm$0.1697)} & \footnotesize{($\pm$0.1206)} & \footnotesize{($\pm$0.1588)} & \footnotesize{($\pm$0.1560)} & \footnotesize{($\pm$0.1237)} \\
    & \multirow{2}{*}{ACCENT} & 0.1889 & \underline{0.5445} & 0.1712 & 0.2089 & \underline{0.5718} & 0.1697 & & 0.2529 & 0.4235 & 0.1671 & 0.3028 & 0.4162 & 0.1603 \\[-0.7ex]
    & & \footnotesize{($\pm$0.2689)} & \footnotesize{($\pm$0.2997)} & \footnotesize{($\pm$0.1270)} & \footnotesize{($\pm$0.2850)} & \footnotesize{($\pm$0.2989)} & \footnotesize{($\pm$0.1277)} & & \footnotesize{($\pm$0.1849)} & \footnotesize{($\pm$0.1733)} & \footnotesize{($\pm$0.1258)} & \footnotesize{($\pm$0.1733)} & \footnotesize{($\pm$0.1441)} & \footnotesize{($\pm$0.1227)} \\
    & \multirow{2}{*}{LXR} & \textbf{0.1331} & 0.4961 & \textbf{0.7040} & \textbf{0.1531} & 0.5327 & \textbf{0.7094} & & \textbf{0.1737} & \underline{0.4446} & \textbf{0.7459} & \textbf{0.2337} & \underline{0.4287} & \textbf{0.7789} \\[-0.7ex]
    & & \footnotesize{($\pm$0.2193)} & \footnotesize{($\pm$0.3266)} & \footnotesize{($\pm$0.1956)} & \footnotesize{($\pm$0.2400)} & \footnotesize{($\pm$0.3225)} & \footnotesize{($\pm$0.1936)} & & \footnotesize{($\pm$0.1667)} & \footnotesize{($\pm$0.1683)} & \footnotesize{($\pm$0.1620)} & \footnotesize{($\pm$0.1671)} & \footnotesize{($\pm$0.1439)} & \footnotesize{($\pm$0.1398)} \\
    \hline
    \multirow{8}{*}{Yahoo} & \multirow{2}{*}{LIME-RS} & \underline{0.1967} & \underline{0.5208} & \underline{0.3286} & \underline{0.2281} & \underline{0.5564} & \underline{0.3307} & & \underline{0.2603} & 0.4218 & \underline{0.3307} & \underline{0.3120} & \underline{0.4366} & \underline{0.3296} \\[-0.7ex]
    & & \footnotesize{($\pm$0.2924)} & \footnotesize{($\pm$0.3326)} & \footnotesize{($\pm$0.1110)} & \footnotesize{($\pm$0.3083)} & \footnotesize{($\pm$0.3332)} & \footnotesize{($\pm$0.1122)} & & \footnotesize{($\pm$0.1824)} & \footnotesize{($\pm$0.1796)} & \footnotesize{($\pm$0.1134)} & \footnotesize{($\pm$0.1557)} & \footnotesize{($\pm$0.1441)} & \footnotesize{($\pm$0.1150)} \\
    & \multirow{2}{*}{SHAP} & 0.2474 & 0.4292 & 0.2867 & 0.2887 & 0.3753 & 0.3016 & & 0.2662 & \textbf{0.4846} & 0.3070 & 0.3314 & 0.4036 & 0.3018 \\[-0.7ex]
    & & \footnotesize{($\pm$0.3005)} & \footnotesize{($\pm$0.3227)} & \footnotesize{($\pm$0.1662)} & \footnotesize{($\pm$0.1794)} & \footnotesize{($\pm$0.1785)} & \footnotesize{($\pm$0.1588)} & & \footnotesize{($\pm$0.3123)} & \footnotesize{($\pm$0.3227)} & \footnotesize{($\pm$0.1696)} & \footnotesize{($\pm$0.1518)} & \footnotesize{($\pm$0.1403)} & \footnotesize{($\pm$0.1404)} \\
    & \multirow{2}{*}{ACCENT} & 0.2471 & 0.4377 & 0.2899 & 0.2791 & 0.4788 & 0.2936 & & 0.3076 & 0.3673 & 0.2813 & 0.3497 & 0.3964 & 0.2884 \\[-0.7ex]
    & & \footnotesize{($\pm$0.3073)} & \footnotesize{($\pm$0.3348)} & \footnotesize{($\pm$0.1800)} & \footnotesize{($\pm$0.3203)} & \footnotesize{($\pm$0.3393)} & \footnotesize{($\pm$0.1803)} & & \footnotesize{($\pm$0.1789)} & \footnotesize{($\pm$0.1822)} & \footnotesize{($\pm$0.1711)} & \footnotesize{($\pm$0.1454)} & \footnotesize{($\pm$0.1491)} & \footnotesize{($\pm$0.1803)} \\
    & \multirow{2}{*}{LXR} & \textbf{0.1534} & \textbf{0.5446} & \textbf{0.7071} & \textbf{0.1868} & \textbf{0.5638} & \textbf{0.7072} & & \textbf{0.2149} & \underline{0.4497} & \textbf{0.7164} & \textbf{0.2802} & \textbf{0.4558} & \textbf{0.7171} \\[-0.7ex]
    & & \footnotesize{($\pm$0.2848)} & \footnotesize{($\pm$0.3361)} & \footnotesize{($\pm$0.2196)} & \footnotesize{($\pm$0.2984)} & \footnotesize{($\pm$0.3372)} & \footnotesize{($\pm$0.2185)} & & \footnotesize{($\pm$0.1853)} & \footnotesize{($\pm$0.1758)} & \footnotesize{($\pm$0.2115)} & \footnotesize{($\pm$0.1566)} & \footnotesize{($\pm$0.1420)} & \footnotesize{($\pm$0.2056)} \\
    \hline
    \end{tabular}
    }
    \label{tab:mf_imp}
\end{table*}

This section presents a comprehensive and reproducible benchmarking study of counterfactual explainers under the framework introduced in Section \ref{sec:benchmark_framework}. We evaluate performance across multiple datasets, examine consistency between item-level and list-level evaluations, and assess robustness across different recommendation architectures. We further analyze stability across item positions, investigate the impact of subgraph-based perturbations for graph explainers, and evaluate computational complexity to assess practical efficiency and scalability.

\subsection{Performance across Different Datasets}

\begin{table*}[!t]
    \centering
    \caption{Performance comparison under the explicit format using MF as the recommendation model (with user interaction vectors as input). S@K denotes PN-S@K, R@K denotes PN-R@K, and \#P@K denotes \#Perturb@K. \textbf{Bold} indicates the best-performing explainer. \underline{Underline} indicates the second-best. $\uparrow$ indicates that higher values correspond to better performance, whereas $\downarrow$ indicates that lower values correspond to better performance. “–” at list-level column denotes that the explainer does not support list-level evaluation.}
    \begin{tabular}{llccccccccccc}
    \hline
    \multirow{2}{*}{\textbf{Dataset}} & \multirow{2}{*}{\textbf{Model}} & \multicolumn{4}{c}{\textbf{Item-level}} & & \multicolumn{6}{c}{\textbf{List-level}} \\\cline{3-6} \cline{8-13}
    & & \textbf{S@3$\uparrow$} & \textbf{\#P@3$\downarrow$} 
    & \textbf{S@5$\uparrow$} & \textbf{\#P@5$\downarrow$} &
    & \textbf{S@3$\uparrow$} & \textbf{R@3$\uparrow$} & \textbf{\#P@3$\downarrow$} 
    & \textbf{S@5$\uparrow$} & \textbf{R@5$\uparrow$} & \textbf{\#P@5$\downarrow$} \\
    \hline
    & \multirow{2}{*}{PRINCE} & 0.2073 & \textbf{0.4287} & \underline{0.2584} & \underline{0.5332} & & \multirow{2}{*}{-} & \multirow{2}{*}{-} & \multirow{2}{*}{-} & \multirow{2}{*}{-} & \multirow{2}{*}{-} & \multirow{2}{*}{-} \\[-0.7ex]
    & & \footnotesize{($\pm$0.4054)} & \footnotesize{($\pm$0.6204)} & \footnotesize{($\pm$0.4378)} & \footnotesize{($\pm$0.5037)} & & & & & & & \\
    \multirow{2}{*}{Amazon} & \multirow{2}{*}{ACCENT} & \underline{0.2600} & \underline{0.5360} & 0.1900 & \textbf{0.4664} & & \multirow{2}{*}{-} & \multirow{2}{*}{-} & \multirow{2}{*}{-} & \multirow{2}{*}{-} & \multirow{2}{*}{-} & \multirow{2}{*}{-} \\[-0.7ex]
    & & \footnotesize{($\pm$0.4386)} & \footnotesize{($\pm$0.4987)} & \footnotesize{($\pm$0.3923)} & \footnotesize{($\pm$0.4989)} & & & & & & & \\
    & \multirow{2}{*}{LXR} & \textbf{0.7420} & 1.8400 & \textbf{0.7032} & 1.8516 & & \textbf{0.7660} & \textbf{0.7587} & \textbf{2.0020} & \textbf{0.7256} & \textbf{0.6970} & \textbf{2.0120} \\[-0.7ex]
    & & \footnotesize{($\pm$0.4375)} & \footnotesize{($\pm$0.7347)} & \footnotesize{($\pm$0.4568)} & \footnotesize{($\pm$0.7494)} & & \footnotesize{($\pm$0.2735)} & \footnotesize{($\pm$0.2865)} & \footnotesize{($\pm$0.7759)} & \footnotesize{($\pm$0.2438)} & \footnotesize{($\pm$0.2720)} & \footnotesize{($\pm$0.7999)} \\
    \hline
    & \multirow{2}{*}{PRINCE} & 0.1400 & \underline{10.2800} & 0.1068 & \underline{8.9552} & & \multirow{2}{*}{-} & \multirow{2}{*}{-} & \multirow{2}{*}{-} & \multirow{2}{*}{-} & \multirow{2}{*}{-} & \multirow{2}{*}{-} \\[-0.7ex]
    & & \footnotesize{($\pm$0.3470)} & \footnotesize{($\pm$35.4574)} & \footnotesize{($\pm$0.3089)} & \footnotesize{($\pm$34.2671)} & & & & & & & \\
    \multirow{2}{*}{ML1M} & \multirow{2}{*}{ACCENT} & \underline{0.1807} & \textbf{0.9993} & \underline{0.1516} & \textbf{0.9980} & & \multirow{2}{*}{-} & \multirow{2}{*}{-} & \multirow{2}{*}{-} & \multirow{2}{*}{-} & \multirow{2}{*}{-} & \multirow{2}{*}{-} \\[-0.7ex]
    & & \footnotesize{($\pm$0.3847)} & \footnotesize{($\pm$0.0258)} & \footnotesize{($\pm$0.3580)} & \footnotesize{($\pm$0.0447)} & & & & & & & \\
    & \multirow{2}{*}{LXR} & \textbf{0.8113} & 11.2080 & \textbf{0.7608} & 10.8464 & & \textbf{0.7933} & \textbf{0.7848} & \textbf{16.3960} & \textbf{0.7476} & \textbf{0.7224} & \textbf{18.7920}\\[-0.7ex]
    & & \footnotesize{($\pm$0.3912)} & \footnotesize{($\pm$5.6471)} & \footnotesize{($\pm$0.4266)} & \footnotesize{($\pm$5.5621)} & & \footnotesize{($\pm$0.2318)} & \footnotesize{($\pm$0.2498)} & \footnotesize{($\pm$6.3497)} & \footnotesize{($\pm$0.1989)} & \footnotesize{($\pm$0.2279)} & \footnotesize{($\pm$7.8908)} \\
    \hline
    & \multirow{2}{*}{PRINCE} & 0.1747 & 3.9987 & 0.1868 & 4.4888 & & \multirow{2}{*}{-} & \multirow{2}{*}{-} & \multirow{2}{*}{-} & \multirow{2}{*}{-} & \multirow{2}{*}{-} & \multirow{2}{*}{-} \\[-0.7ex]
    & & \footnotesize{($\pm$0.3797)} & \footnotesize{($\pm$12.5242)} & \footnotesize{($\pm$0.3898)} & \footnotesize{($\pm$13.4623)} & & & & & & & \\
    \multirow{2}{*}{Yahoo} & \multirow{2}{*}{ACCENT} & \underline{0.2673} & \textbf{0.9160} & \underline{0.2284} & \textbf{0.8984} & & \multirow{2}{*}{-} & \multirow{2}{*}{-} & \multirow{2}{*}{-} & \multirow{2}{*}{-} & \multirow{2}{*}{-} & \multirow{2}{*}{-} \\[-0.7ex]
    & & \footnotesize{($\pm$0.4426)} & \footnotesize{($\pm$0.2774)} & \footnotesize{($\pm$0.4198)} & \footnotesize{($\pm$0.3021)} & & & & & & & \\
    & \multirow{2}{*}{LXR} & \textbf{0.7087} & \underline{2.4853} & \textbf{0.6540} & \underline{2.4604} & & \textbf{0.6927} & \textbf{0.6777} & \textbf{4.6820} & \textbf{0.6476} & \textbf{0.6124} & \textbf{5.6980} \\[-0.7ex]
    & & \footnotesize{($\pm$0.4544)} & \footnotesize{($\pm$1.4006)} & \footnotesize{($\pm$0.4757)} & \footnotesize{($\pm$1.4872)} & & \footnotesize{($\pm$0.2724)} & \footnotesize{($\pm$0.2907)} & \footnotesize{($\pm$2.0892)} & \footnotesize{($\pm$0.2444)} & \footnotesize{($\pm$0.2683)} & \footnotesize{($\pm$2.6644)} \\
    \hline
    \end{tabular}
    \label{tab:mf_exp}
\end{table*}

\begin{table*}[!t]
    \centering
    \setlength{\tabcolsep}{4pt}
    \caption{Performance comparison under explicit setting with LightGCN as recommendation model (user-item graph as input). S@K represents PN-S@K, R@K represents PN-R@K, \#P@K represents \#Perturb at K. \textbf{Bold} represents the best explainer. \underline{Underline} represents the second best explainer. $\uparrow$ indicates that higher values correspond to better performance, whereas $\downarrow$ indicates that lower values correspond to better performance. "-" represents out-of-memory.}
    \resizebox{\textwidth}{!}{
    \begin{tabular}{llccccccccccc}
    \hline
    \multirow{2}{*}{\textbf{Dataset}} & \multirow{2}{*}{\textbf{Model}} & \multicolumn{4}{c}{\textbf{Item-level}} & & \multicolumn{6}{c}{\textbf{List-level}} \\\cline{3-6} \cline{8-13}
    & & \textbf{S@3$\uparrow$} & \textbf{\#P@3$\downarrow$} 
    & \textbf{S@5$\uparrow$} & \textbf{\#P@5$\downarrow$} & 
    & \textbf{S@3$\uparrow$} & \textbf{R@3$\uparrow$} & \textbf{\#P@3$\downarrow$} 
    & \textbf{S@5$\uparrow$} & \textbf{R@5$\uparrow$} & \textbf{\#P@5$\downarrow$} \\
    \hline
    & \multirow{2}{*}{GREASE} & \textbf{0.8513} & 4.99 & \textbf{0.8392} & 9.29 & & \textbf{0.9113} & \textbf{0.9341} & 13.46 & \textbf{0.8012} & \textbf{0.7977} & 85.12 \\[-0.7ex] 
    & & \footnotesize{($\pm$0.1396)} & \footnotesize{($\pm$2.37)} & \footnotesize{($\pm$0.1105)} & \footnotesize{($\pm$2.40)} & & \footnotesize{($\pm$0.3034)} & \footnotesize{($\pm$0.3180)} & \footnotesize{($\pm$6.80)} & \footnotesize{($\pm$0.2809)} & \footnotesize{($\pm$0.3054)} & \footnotesize{($\pm$17.05)} \\
    & \multirow{2}{*}{CF-GNNExplainer} & 0.4467 & \textbf{1.12} & 0.3064 & \textbf{0.77} & & 0.3813 & 0.3166 & \textbf{1.19} & 0.2692 & 0.2059 & \textbf{1.14} \\[-0.7ex]
    & & \footnotesize{($\pm$0.4971)} & \footnotesize{($\pm$1.56)} & \footnotesize{($\pm$0.4610)} & \footnotesize{($\pm$1.33)} & & \footnotesize{($\pm$0.2085)} & \footnotesize{($\pm$0.2120)} & \footnotesize{($\pm$0.83)} & \footnotesize{($\pm$0.1623)} & \footnotesize{($\pm$0.1468)} & \footnotesize{($\pm$0.72)} \\
    & \multirow{2}{*}{CF$^2$} & 0.5422 & 24.87 & 0.4247 & 24.79 & & 0.5360 & 0.5015 & 24.94 & 0.4196 & 0.3716 & 25.49 \\[-0.7ex]
    & & \footnotesize{($\pm$0.2543)} & \footnotesize{($\pm$21.28)} & \footnotesize{($\pm$0.2121)} & \footnotesize{($\pm$20.71)} & & \footnotesize{($\pm$0.2334)} & \footnotesize{($\pm$0.2549)} & \footnotesize{($\pm$21.15)} & \footnotesize{($\pm$0.2060)} & \footnotesize{($\pm$0.2181)} & \footnotesize{($\pm$20.33)} \\
    \multirow{2}{*}{Amazon} & \multirow{2}{*}{CLEAR} & \underline{0.6340} & 9.25 & \underline{0.6876} & 9.39 & & 0.3660 & 0.3190 & 9.25 & 0.3124 & 0.2549 & 9.39 \\[-0.7ex]
    & & \footnotesize{($\pm$0.4230)} & \footnotesize{($\pm$7.52)} & \footnotesize{($\pm$0.4271)} & \footnotesize{($\pm$8.46)} & & \footnotesize{($\pm$0.1658)} & \footnotesize{($\pm$0.1208)} & \footnotesize{($\pm$7.52)} & \footnotesize{($\pm$0.2494)} & \footnotesize{($\pm$0.2156)} & \footnotesize{($\pm$7.96)} \\
    & \multirow{2}{*}{UNR-Explainer} & 0.5836 & \underline{1.85} & 0.4833 & \underline{2.05} & & \underline{0.6380} & 0.5681 & \underline{2.85} & \underline{0.5244} & 0.4373 & \underline{2.96} \\[-0.7ex]
    & & \footnotesize{($\pm$0.2902)} & \footnotesize{($\pm$1.14)} & \footnotesize{($\pm$0.2394)} & \footnotesize{($\pm$1.18)} & & \footnotesize{($\pm$0.2613)} & \footnotesize{($\pm$0.2917)} & \footnotesize{($\pm$2.11)} & \footnotesize{($\pm$0.2197)} & \footnotesize{($\pm$0.2326)} & \footnotesize{($\pm$1.88)} \\
    & \multirow{2}{*}{C2Explainer (-add)} & 0.4840 & 91.08 & 0.4136 & 87.77 & & 0.4840 & 0.4459 & 126.75 & 0.4136 & 0.3533 & 159.75 \\[-0.7ex]
    & & \footnotesize{($\pm$0.4652)} & \footnotesize{($\pm$60.48)} & \footnotesize{($\pm$0.4854)} & \footnotesize{($\pm$63.07)} & & \footnotesize{($\pm$0.2494)} & \footnotesize{($\pm$0.2694)} & \footnotesize{($\pm$67.61)} & \footnotesize{($\pm$0.2304)} & \footnotesize{($\pm$0.2468)} & \footnotesize{($\pm$94.93)} \\
    & \multirow{2}{*}{C2Explainer} & 0.6100 & 157.39 & 0.5220 & 111.33 & & 0.6100 & \underline{0.5838} & 117.39 & 0.5220 & \underline{0.4749} & 170.33 \\[-0.7ex]   
    & & \footnotesize{($\pm$0.4482)} & \footnotesize{($\pm$84.04)} & \footnotesize{($\pm$0.4128)} & \footnotesize{($\pm$80.60)} & & \footnotesize{($\pm$0.2682)} & \footnotesize{($\pm$0.2681)} & \footnotesize{($\pm$56.78)} & \footnotesize{($\pm$0.2603)} & \footnotesize{($\pm$0.2381)} & \footnotesize{($\pm$81.80)} \\
    
    \hline
    & \multirow{2}{*}{GREASE} & 0.1867 & 32.77 & 0.1240 & 33.84 & & 0.3200 & 0.2778 & 61.08 & 0.2560 & 0.2082 & 72.56\\[-0.7ex]
    & & \footnotesize{($\pm$0.1352)} & \footnotesize{($\pm$10.98)} & \footnotesize{($\pm$0.1176)} & \footnotesize{($\pm$11.31)} & & \footnotesize{($\pm$0.1622)} & \footnotesize{($\pm$0.1591)} & \footnotesize{($\pm$21.02)} & \footnotesize{($\pm$0.1389)} & \footnotesize{($\pm$0.1194)} & \footnotesize{($\pm$28.95)} \\
    & \multirow{2}{*}{CF-GNNExplainer} & 0.3667 & \underline{4.58} & 0.2200 & \underline{1.96} & & 0.3000 & 0.2132 & \underline{5.40} & 0.2000 & 0.1350 & \underline{4.60} \\[-0.7ex]
    & & \footnotesize{($\pm$0.4819)} & \footnotesize{($\pm$7.76)} & \footnotesize{($\pm$0.4142)} & \footnotesize{($\pm$4.36)} & &  \footnotesize{($\pm$0.1000)} & \footnotesize{($\pm$0.0719)} & \footnotesize{($\pm$7.34)} & \footnotesize{($\pm$0.0730)} & \footnotesize{($\pm$0.0548)} & \footnotesize{($\pm$3.98)} \\
    & \multirow{2}{*}{CF$^2$} &  0.5453 & 62125.14 & 0.5356 & 66373.28 & & 0.5127 & 0.5991 & 52522.40 & 0.4192 & 0.4923 & 61031.18 \\[-0.7ex]
    & & \footnotesize{($\pm$0.2243)} & \footnotesize{($\pm$27900.03)} & \footnotesize{($\pm$0.0631)} & \footnotesize{($\pm$30079.05)} & & \footnotesize{($\pm$0.2232)} & \footnotesize{($\pm$0.1820)} & \footnotesize{($\pm$27515.74)} & \footnotesize{($\pm$0.1950)} & \footnotesize{($\pm$0.1626)} & \footnotesize{($\pm$28213.56)} \\
    ML1M & CLEAR & - & - & - & - & & - & - & - & - & - & - \\
    & \multirow{2}{*}{UNR-Explainer} & 0.1911 & \textbf{1.22} & 0.1308 & \textbf{1.20} & & 0.1633 & 0.1174 & \textbf{1.13} & 0.1140 & 0.0771 & \textbf{1.07}\\[-0.7ex]
    & & \footnotesize{($\pm$0.1820)} & \footnotesize{($\pm$0.77)} & \footnotesize{($\pm$0.1372)} & \footnotesize{($\pm$0.69)} & & \footnotesize{($\pm$0.1666)} & \footnotesize{($\pm$0.1204)} & \footnotesize{($\pm$0.72)} & \footnotesize{($\pm$0.1304)} & \footnotesize{($\pm$0.0950)} & \footnotesize{($\pm$0.41)} \\
    & \multirow{2}{*}{C2Explainer (-add)} & \underline{0.8867} & 9785.37 & \underline{0.8188} & 9021.06 & & \underline{0.8867} & \underline{0.9043} & 9785.37 & \underline{0.8188} & \underline{0.8465} & 9021.06 \\[-0.7ex]
    & & \footnotesize{($\pm$0.3210)} & \footnotesize{($\pm$4437.54)} & \footnotesize{($\pm$0.2713)} & \footnotesize{($\pm$5371.02)} & & \footnotesize{($\pm$0.2230)} & \footnotesize{($\pm$0.2354)} & \footnotesize{($\pm$3662.02)} & \footnotesize{($\pm$0.1658)} & \footnotesize{($\pm$0.1621)} & \footnotesize{($\pm$4594.91)} \\
    & \multirow{2}{*}{C2Explainer} & \textbf{0.9753} & 222184.59 & \textbf{0.9336} & 221160.92 & & \textbf{0.9753} & \textbf{0.9807} & 222184.59 & \textbf{0.9336} & \textbf{0.9520} & 221160.92 \\[-0.7ex]
    & & \footnotesize{($\pm$0.2764)} & \footnotesize{($\pm$47467.78)} & \footnotesize{($\pm$0.2713)} & \footnotesize{($\pm$52349.27)} & & \footnotesize{($\pm$0.1443)} & \footnotesize{($\pm$0.1076)} & \footnotesize{($\pm$44659.65)} & \footnotesize{($\pm$0.0995)} & \footnotesize{($\pm$0.0742)} & \footnotesize{($\pm$54035.91)} \\
    \hline
    & \multirow{2}{*}{GREASE} & \underline{0.2667} & \underline{24.28} & 0.2040 & \underline{25.45} & & \underline{0.4400} & \underline{0.4108} & \underline{52.36} & \underline{0.3080} & \underline{0.2617} & \underline{61.06}\\[-0.7ex]
    & & \footnotesize{($\pm$0.1591)} & \footnotesize{($\pm$8.33)} & \footnotesize{($\pm$0.1429)} & \footnotesize{($\pm$9.15)} & & \footnotesize{($\pm$0.2365)} & \footnotesize{($\pm$0.2291)} & \footnotesize{($\pm$17.71)} & \footnotesize{($\pm$0.1541)} & \footnotesize{($\pm$0.1279)} & \footnotesize{($\pm$20.07)} \\
    & CF-GNNExplainer & - & - & - & - & & - & - & - & - & - & -  \\
    & \multirow{2}{*}{CF$^2$} & \textbf{0.5311} & 19813.02 & \textbf{0.4780} & 18522.44 & & \textbf{0.4800} & \textbf{0.5301} & 16218.90 & \textbf{0.3927} & \textbf{0.4205} & 16276.30 \\[-0.7ex]
    & & \footnotesize{($\pm$0.2407)} & \footnotesize{($\pm$13645.52)} & \footnotesize{($\pm$0.2528)} & \footnotesize{($\pm$13624.24)} & & \footnotesize{($\pm$0.2083)} & \footnotesize{($\pm$0.2057)} & \footnotesize{($\pm$12878.67)} & \footnotesize{($\pm$0.2115)} & \footnotesize{($\pm$0.2031)} & \footnotesize{($\pm$12890.96)} \\
    Yahoo & CLEAR & - & - & - & - & & - & - & - & - & - & -\\
    & \multirow{2}{*}{UNR-Explainer} & 0.2367 & \textbf{1.50} & \underline{0.2080} & \textbf{1.66} & & 0.1967 & 0.1432 & \textbf{1.41} & 0.1740 & 0.1225 & \textbf{1.54} \\[-0.7ex]
    & & \footnotesize{($\pm$0.4250)} & \footnotesize{($\pm$0.99)} & \footnotesize{($\pm$0.4059)} & \footnotesize{($\pm$1.15)} & & \footnotesize{($\pm$0.1949)} & \footnotesize{($\pm$0.1506)} & \footnotesize{($\pm$0.88)} & \footnotesize{($\pm$0.1514)} & \footnotesize{($\pm$0.1193)} & \footnotesize{($\pm$1.14)} \\
    & C2Explainer (-add) & - & - & - & - & & - & - & - & - & - & - \\
    & C2Explainer & - & - & - & - & & - & - & - & - & - & - \\
    \hline
    
    \end{tabular}
    }
    \label{tab:lightgcn_exp}
\end{table*}

We first evaluate counterfactual explainers across multiple datasets. Due to space constraints, we focus on two representative models: MF for user-vector-based recommenders and LightGCN for graph-based recommenders. Results for MF under implicit and explicit settings are shown in Tables \ref{tab:mf_imp} and \ref{tab:mf_exp}, while LightGCN results are reported in Table \ref{tab:lightgcn_exp}. For graph-based recommenders, experiments are conducted only under the explicit format, as existing explainers do not support implicit explanations. Consistency across additional recommendation models is discussed in Section \ref{subsec:different_rec_model}.

Under the implicit format with MF, LXR exhibits a clear advantage in both effectiveness and sparsity. In terms of effectiveness, across different datasets and top-$K$ evaluation, it achieves the lowest POS-P score and the highest NEG-P score in most of the cases, indicating superior capability of the learning-based paradigm for CE generation. These findings are consistent with the results reported in the original paper, further validating the reproducibility in our evaluation process. In terms of sparsity, LXR consistently outperforms the competing methods, often by a substantial margin, particularly on the Amazon and ML1M datasets, where it obtains the highest Gini Index values. This improvement can be attributed to the additional $L_1$ regularization imposed during mask optimization, which explicitly encourages sparse and more concise explanations.

Under the explicit format, across both MF and LightGCN, a consistent observation is that no single explainer uniformly dominates all evaluation metrics. In particular, a trade-off emerges between effectiveness and explanation sparsity at both the item and list levels: methods achieving higher effectiveness generally produce less sparse explanations (reflected by a larger number of required perturbations), whereas models generating sparser explanations tend to exhibit lower effectiveness. However, this trade-off does not hold universally. For instance, on the Amazon dataset under item-level evaluation, UNR-Explainer generates more effective counterfactual explanations while requiring fewer than three perturbations on average, whereas CF$^2$ and C2Explainer (-add) require approximately 25 and 91 perturbations, respectively, to achieve lower effectiveness. \textit{Regarding MF}, LXR continues to achieve superior item-level effectiveness compared to the other CE models similar to under the implicit setting, but requires a larger number of perturbations, differing from the pattern observed in the implicit case. This discrepancy may stem from differences in interaction selection strategies: ACCENT incrementally evaluates changes in model predictions through score gaps during the search process, thereby implicitly controlling explanation size, whereas LXR relies on a fixed threshold over item importance scores, which does not explicitly optimize the number of interaction removals. \textit{Regarding LightGCN}, the number of perturbed edges varies notably across datasets, largely due to differences in interaction density. On Amazon, which contains fewer interactions, all explainers require relatively small perturbation sets. In contrast, ML1M and Yahoo, particularly the denser ML1M dataset, lead to substantially larger perturbation sizes. In terms of effectiveness, GREASE, a counterfactual explainer specifically designed for recommender systems, achieves strong performance on Amazon but degrades substantially on ML1M and Yahoo, where it attains the lowest effectiveness in the item-level evaluation. CF$^2$ and C2Explainer achieve the highest effectiveness overall; however, they require extremely large perturbation sets (reaching tens of thousands of interactions on ML1M and Yahoo), thereby limiting interpretability. Conversely, CF-GNNExplainer and UNR-Explainer generate much smaller perturbation sets, although with more modest effectiveness. These differences stem from their optimization strategies: CF$^2$ and C2Explainer minimize a target loss starting from large edge sets and progressively reducing them, whereas CF-GNNExplainer maximizes an objective with a conditional mask selection criteria during training, and UNR-Explainer adopts a search-based strategy, both of which expands perturbations from small to larger sets. CLEAR performs competitively on Amazon but scales poorly to larger datasets due to the need to introduce a causal variable for each interaction, resulting in substantial memory overhead.

\subsection{Explainer Consistency across Different Evaluation Levels}

List-level evaluation provides a complementary perspective on the CE quality by measuring their impact on the entire ranked list rather than on a single target item. Since item-level and list-level evaluations reflect different aspects of model behavior, strong performance under one level does not necessarily generalize to the other. Therefore, examining consistency under different evaluation levels is essential to assess the stability, reliability, and generalization of counterfactual explainers across evaluation protocols.

As shown in Table \ref{tab:mf_imp} and Table \ref{tab:mf_exp}, the explainers exhibit largely consistent trends across item-level and list-level evaluations. For the MF model under the implicit format, LXR achieves the highest effectiveness on Amazon and Yahoo at both evaluation levels, except for NEG-P@3 at the list-level evaluation on Yahoo, where SHAP exhibits the best performance. A similar pattern is observed on ML1M, where LXR leads in POS-P, while LIME-RS attains the best performance in NEG-P. In terms of sparsity, LXR consistently outperforms all competing methods across the three datasets at both evaluation levels. Under the explicit format, the list-level comparison is restricted to LXR, as PRINCE and ACCENT do not support list-level counterfactual generation under this setting. Notably, LXR maintains highly consistent scores between item-level and list-level evaluations across all datasets, further demonstrating its robustness across evaluation formats.

A comparable degree of consistency is observed for graph-based explainers on LightGCN (illustrated in Table \ref{tab:lightgcn_exp}), where most methods maintain similar relative effectiveness across the two evaluation levels. However, an exception arises on the Amazon dataset: CLEAR ranks second in item-level effectiveness but drops to the second-worst position at the list level and performs worst on PN-S@3. Whether this discrepancy reflects a systematic limitation remains unclear, as experiments on ML1M and Yahoo could not be completed due to out-of-memory constraints.

\subsection{Explainer Consistency across Different Recommender Architectures}
\label{subsec:different_rec_model}

\begin{figure}
    \centering
    \includegraphics[width=\linewidth]{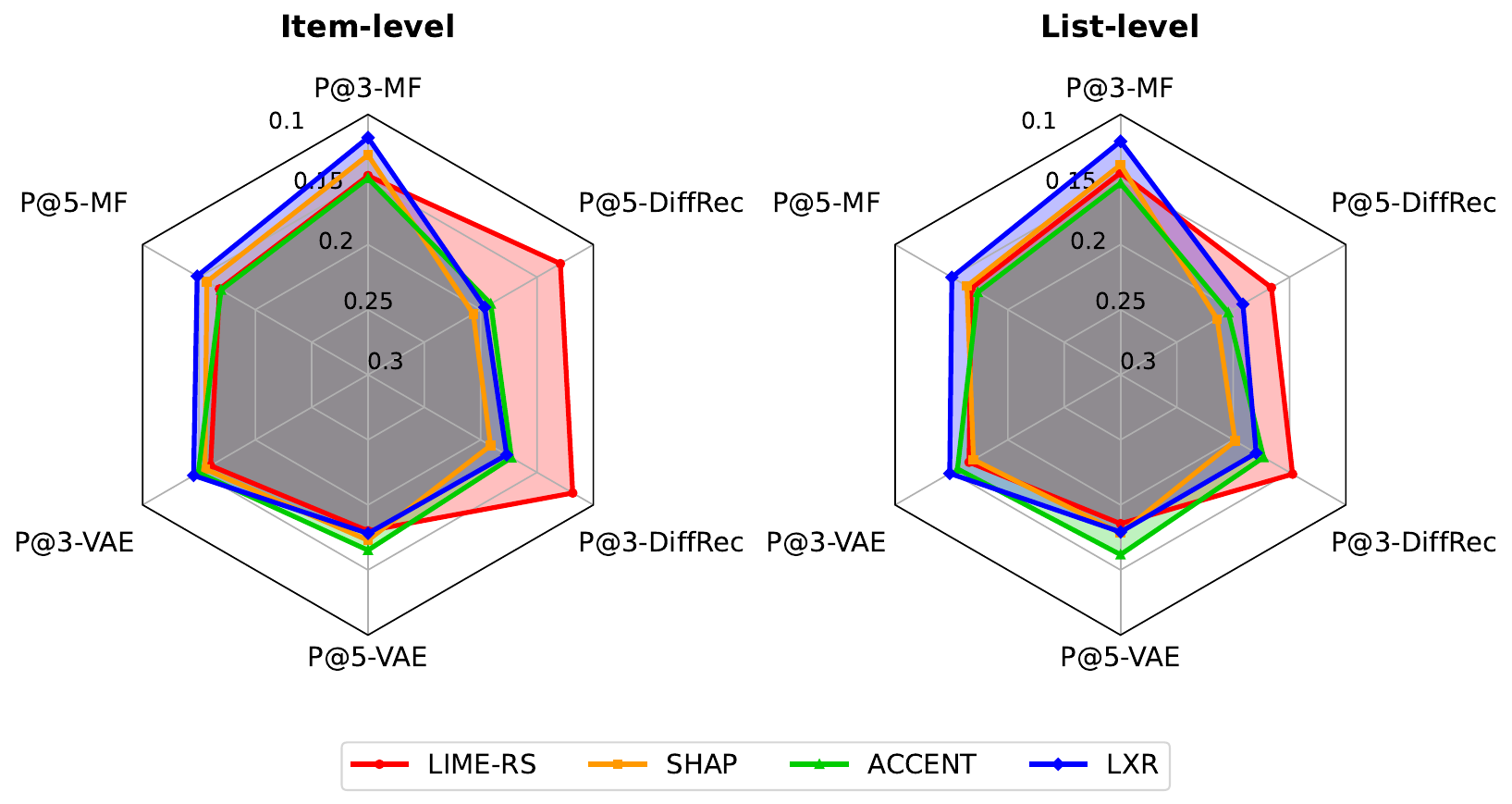}
    \caption{Performance across different user-vector-based recommendation models on the Amazon dataset under implicit format. P denotes POS-P.}
    \label{fig:diff_uv_rec_model}
\end{figure}

First, we evaluate explainer robustness across recommendation architectures under multiple user-vector-based models on Amazon (Figure \ref{fig:diff_uv_rec_model}). While performance differences are relatively minor for MF and VAE, greater variation is observed for DiffRec. In particular, although LXR demonstrates strong effectiveness in terms of POS-P under MF and VAE, LIME-RS achieves the highest performance on DiffRec, with a notably large margin. These findings suggest limited cross-recommender stability among the explainers.

\begin{figure}[t!]
    \centering
    \includegraphics[width=\linewidth]{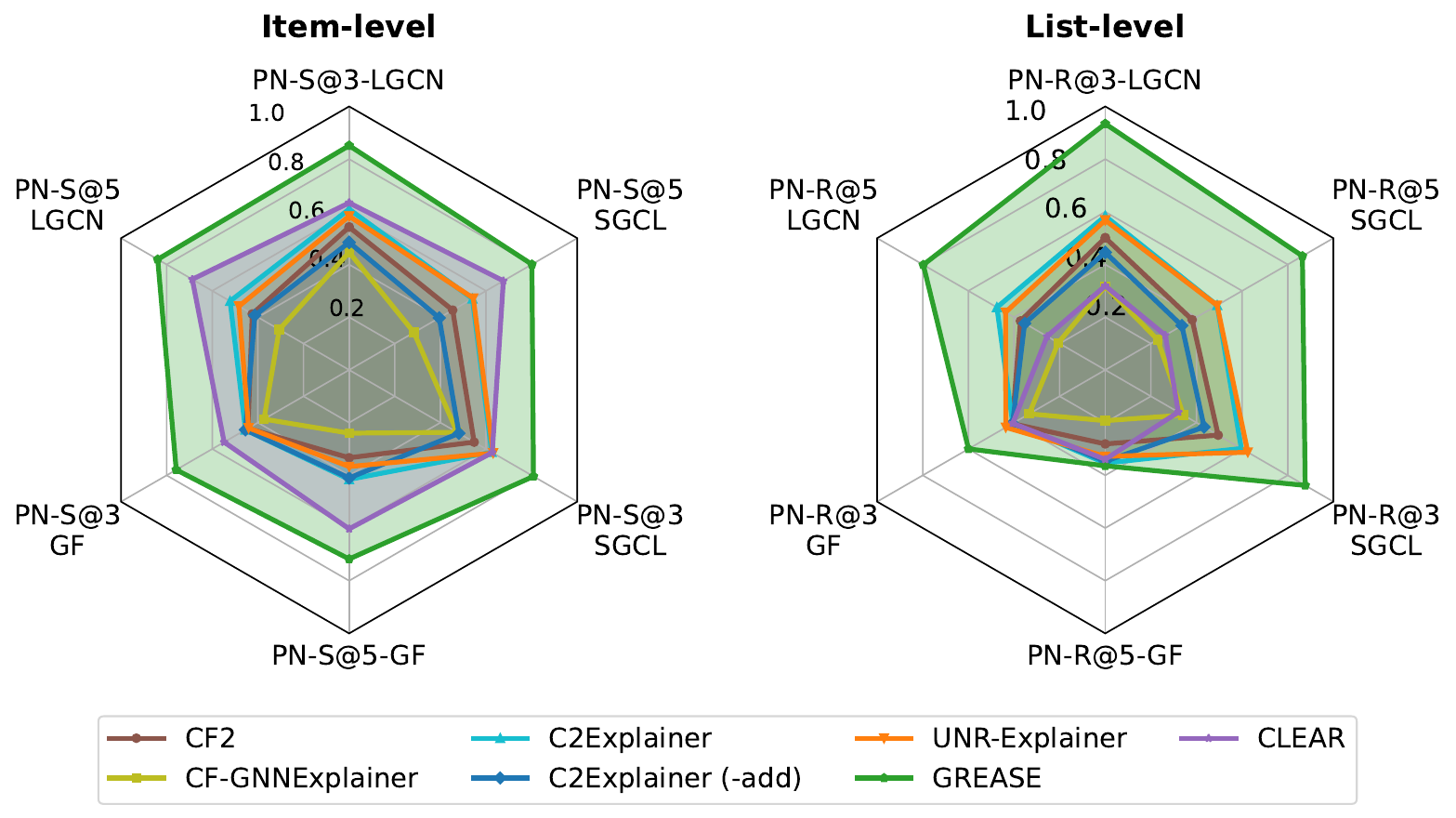}
    \caption{Performance across different graph-based recommendation models on the Amazon dataset under explicit format. LGCN represents LightGCN, GF represents GFormer, and SGCL represents SimGCL.}
    \label{fig:diff_graph_rec_model}
\end{figure}

Figure \ref{fig:diff_graph_rec_model} illustrates the effectiveness of explainers across three graph-based recommenders on the Amazon dataset. The radar plots indicate a generally stable performance pattern across metrics within the same evaluation level, despite changes in the underlying recommender. At the item level, GREASE and CLEAR consistently rank first and second, respectively, while the remaining explainers follow with relatively small performance gaps. At the list level, GREASE continues to outperform other methods across recommenders, with C2Explainer and UNR-Explainer exhibiting stable performance as the next best approaches. An interesting observation arises for GFormer under list-level evaluation, where the explainers yield nearly similar results, suggesting reduced sensitivity to the choice of explanation method in this setting.

\subsection{Stability across Different Item Position in Item-level Evaluation}
In item-level evaluation, we assess CEs by checking whether a target item is removed from the top-$K$ list after perturbation. A natural assumption is that lower-ranked items (closer to position $K$) are easier to be displaced than top-ranked ones due to smaller score gaps. To verify whether this holds across models, we evaluate explainer stability across different item positions in the top-$K$ list.

\begin{figure}[!t]
    \centering
    \includegraphics[width=\linewidth]{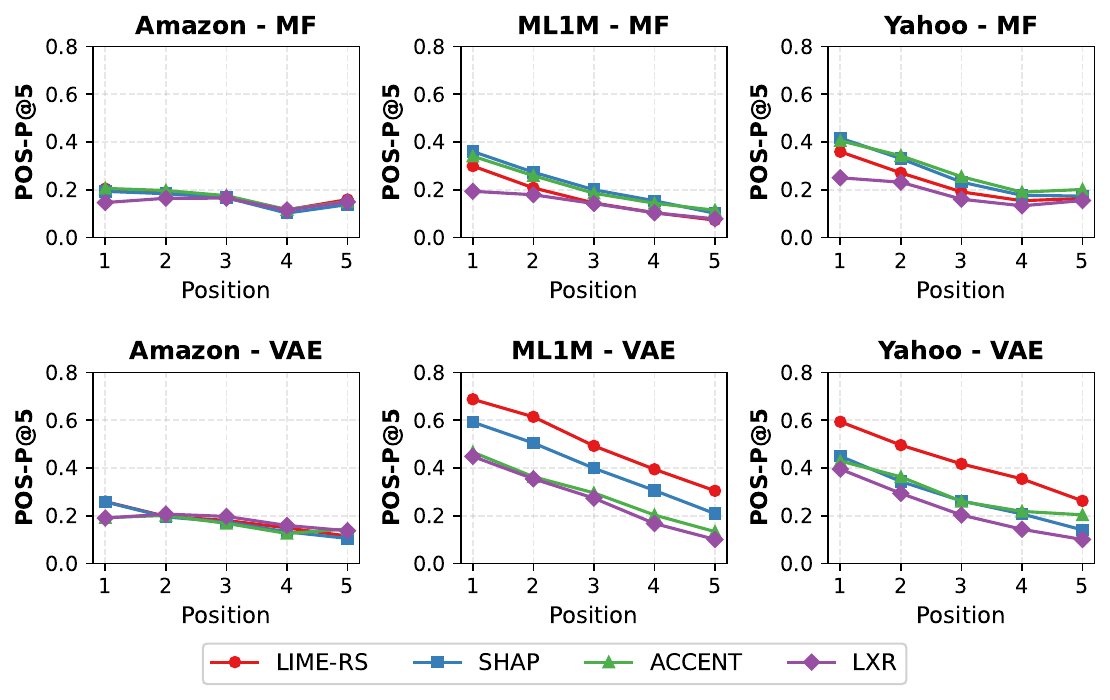}
    \caption{User-vector-based explainer performance under item-level evaluation w.r.t.   item position in top-5 lists.}
    \label{fig:diff_top_k_uv}
\end{figure}

We first analyze explainers applied to user-vector recommenders. Figure \ref{fig:diff_top_k_uv} reports the results under the implicit format for items at different positions within the top-5 list. Across methods, POS-P@5 consistently decreases from position 1 to 5, indicating that generating counterfactual explanations becomes easier for lower-ranked items. This trend is more pronounced on ML1M and Yahoo, and less distinct on Amazon. A plausible explanation is that, in larger datasets such as ML1M and Yahoo, the larger search space and weaker marginal influence of individual interactions make highly ranked items more difficult to perturb, whereas the smaller scale of Amazon reduces such positional disparities.

\begin{figure}[!t]
    \centering
    \includegraphics[width=\linewidth]{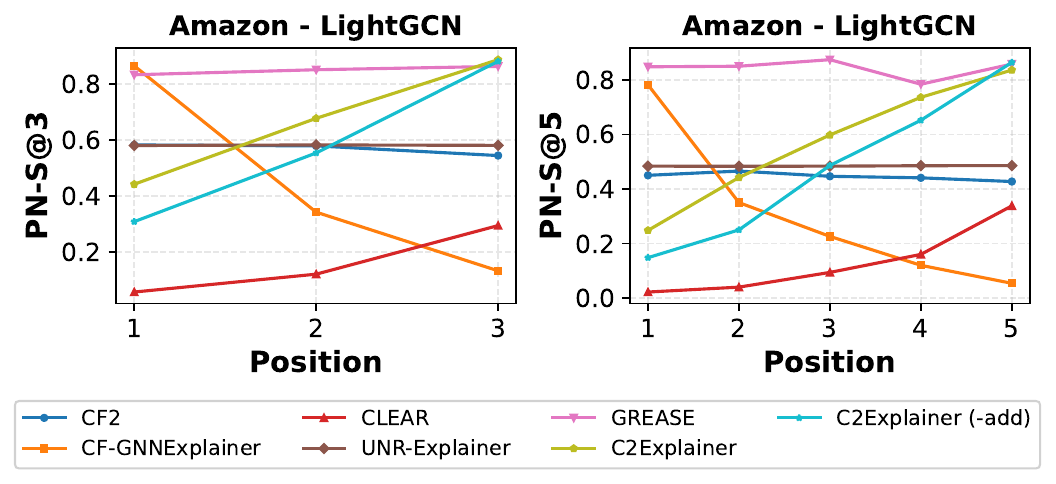}
    \caption{Graph-based explainer performance under item-level evaluation w.r.t. items at different position in top-3 lists (\textit{the left image}) and top-5 lists (\textit{the right image}).}
    \label{fig:diff_top_k_graph}
\end{figure}

We next examine explainers for graph-based recommenders. Figure \ref{fig:diff_top_k_graph} presents the results under the explicit setting for items at different positions in the top-3 and top-5 lists. Positional stability varies across methods. CLEAR and C2Explainer follow the expected trend, showing improved effectiveness for lower-ranked items, whereas CF-GNNExplainer does not exhibit this pattern. In contrast, GREASE, CF$^2$, and UNR-Explainer maintain relatively stable performance across positions, indicating lower sensitivity to the ranking of the target item.


\subsection{Effect of Subgraph versus Full Graph Perturbation on Graph-based Explainers}

In GNN-based explainers, particularly for graph-based recommendation systems, it is common to reduce the perturbation scope by extracting a $k$-hop subgraph around the target node \cite{lucic2022cf,tan2022learning} or by combining the $k$-hop neighborhoods of the user and item nodes \cite{chen2022grease,chen2025joint}. This practice is motivated by message passing of GNN, where a node’s representation depends only on its $k$-hop neighborhood. 
However, in recommendation setting, perturbing a single edge may influence multiple node representations and indirectly affect the top-$K$ list. Therefore, it is necessary to investigate whether restricting explanations to $k$-hop subgraphs is appropriate in this scenario.

\begin{figure}[!t]
    \centering
    \includegraphics[width=\linewidth]{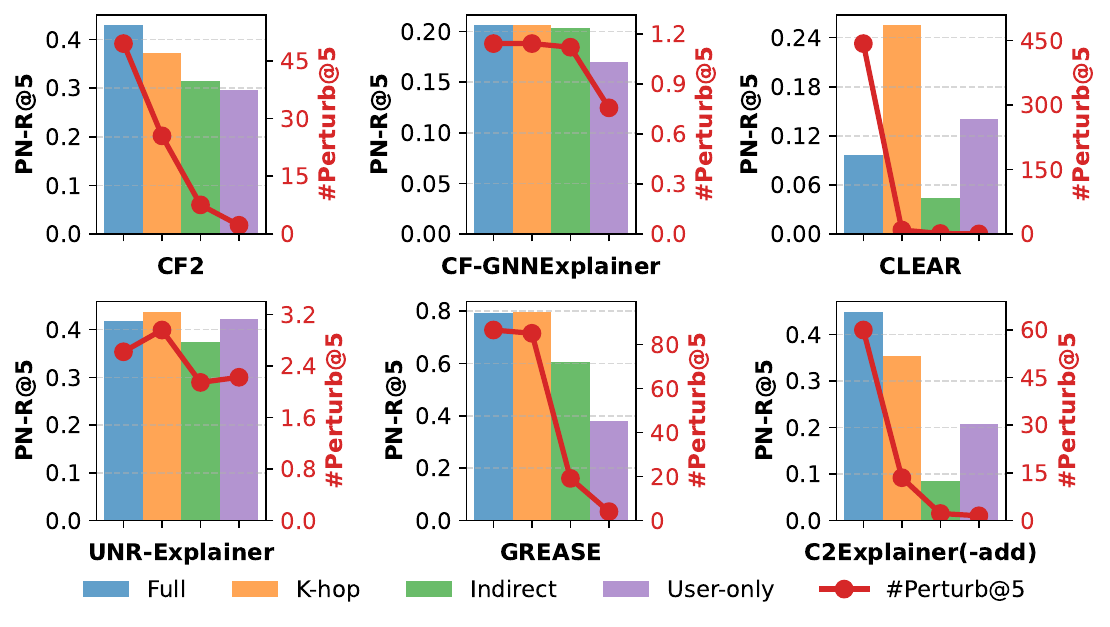}
    \caption{Explainer performance under varying graph perturbation scopes on the Amazon dataset (list-level evaluation) with LightGCN as recommender.}
    \label{fig:graph_perturb_list}
\end{figure}

\begin{figure}[!t]
    \centering
    \includegraphics[width=\linewidth]{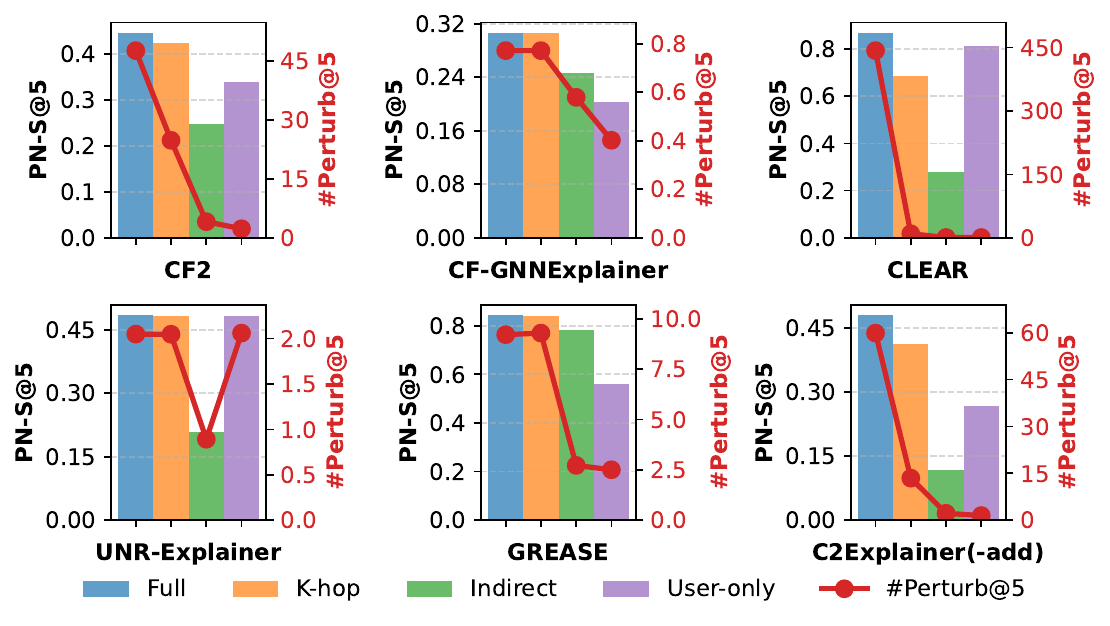}
    \caption{Explainer performance under varying graph perturbation scopes on the Amazon dataset (item-level evaluation) with LightGCN as recommender.}
    \label{fig:graph_perturb_item}
\end{figure}

To investigate this, we design four scenarios with progressively restricted perturbation scopes: 
(1) \textit{Full}, using the entire graph; (2) \textit{$k$-hop}, constructing a subgraph from the $k$-hop neighborhoods of the user and target item (the default for methods such as GREASE, CF-GNNExplainer, and CF$^2$); (3) \textit{Indirect}, retaining only the $k$-hop paths between the user and the target item; and (4) \textit{User-only}, limiting perturbations to the user’s interaction vector. As shown in Figure ~\ref{fig:graph_perturb_list} and Figure ~\ref{fig:graph_perturb_item}, using $k$-hop subgraphs reduces performance for CF$^2$ and C2Explainer compared to the full graph, but yields comparable results for CF-GNNExplainer, GREASE, and UNR-Explainer. For CLEAR and UNR-Explainer under list-level evaluation, subgraph restriction even leads to notable improvements, suggesting that limiting the perturbation scope can effectively reduce the search space while preserving explanation quality. In contrast, restricting perturbations to $k$-hop paths provides no consistent gains and often underperforms the user-only setting, likely because path extraction removes direct links to influential items while retaining indirect connections with limited impact on the target ranking.

\subsection{Computational Complexity}
Finally, we assess the computational complexity of each explainer by measuring runtime across datasets (Table \ref{tab:comp2}). In general, list-level CE generation is more time-consuming than item-level generation, as evaluating changes over an entire ranked list requires more comparisons than evaluating a single item. Pretraining-based methods, such as LXR and CLEAR, exhibit minimal inference time across all datasets, demonstrating the practicality of shifting computational cost to the offline training stage. However, pretraining itself is computationally demanding; notably, CLEAR encounters out-of-memory issues on ML1M and Yahoo, revealing scalability limitations on larger datasets.

Among user-vector-based explainers, SHAP incurs substantially longer runtimes on larger datasets (i.e., ML1M and Yahoo) due to the exponential complexity of computing Shapley values. In contrast, LIME-RS trains a local surrogate model within the restricted user interaction space, ensuring efficient computation and the fastest inference time. PRINCE and ACCENT adopt search-based strategies and are therefore slower than LIME-RS; nevertheless, their stopping mechanisms help maintain moderate runtimes in practice.

For graph-based explainers, both optimization-based methods (GREASE, CF-GNNExplainer, CF$^2$, and C2Explainer) and the tree-search-based UNR-Explainer run efficiently on the smaller Amazon dataset (except GREASE), but incur substantially longer runtimes on ML1M and Yahoo. This increase arises from repeated optimization steps, each requiring model queries to evaluate counterfactual effectiveness, which becomes costly on larger datasets. Notably, UNR-Explainer achieves faster execution than the other methods, likely because of its search-based strategy that avoids expensive gradient-based optimization over large matrices.

\begin{table}[!t]
    \centering
    \caption{Comparison on average inference time (\textit{seconds}) to generate an explanation (top-5 recommendation). For graph-based explainers that do not require pretraining, the results are reported after 100 optimization/search steps. OOM represents out-of-memory. “-” at list-level column denotes that the explainer does not support list-level evaluation.}
    \setlength{\tabcolsep}{3pt}
    \resizebox{\linewidth}{!}{
    \begin{tabular}{lccccccc}
    \hline
       \multirow{2}{*}{\textbf{Explainer}} & \multicolumn{3}{c}{\textbf{Item-level}} & & \multicolumn{3}{c}{\textbf{List-level}} \\\cline{2-4} \cline{6-8}
        & \textbf{Amazon} & \textbf{ML1M} & \textbf{Yahoo} & & \textbf{Amazon} & \textbf{ML1M} & \textbf{Yahoo}\\
       \hline
       \multicolumn{5}{l}{\textit{User vector (MF as recommender)}} \\
       \hline
       LIME-RS & 0.016 & 0.026 & 0.028 & & 0.025 & 0.034 & 0.038 \\
       SHAP & 0.111 & 0.732 & 1.254 & & 0.050 & 1.559 & 2.091 \\
        PRINCE & 0.161 & 0.311 & 0.421 & & - & - & - \\
       ACCENT & 0.205 & 0.304 & 0.213 & & 1.759 & 1.869 & 1.852 \\
       LXR \textit{(pre-train)} & 0.001 & 0.002 & 0.002 & & 0.001 & 0.003 & 0.004 \\
       \hline
       \multicolumn{5}{l}{\textit{Graph (LightGCN as recommender)}} \\
       \hline
       GREASE & 6.717 & 98.212 & 348.751 & & 7.506 & 137.412 & 598.311 \\
       CF-GNNExplainer & 5.627 & 216.359 & OOM & & 7.023 & 265.413 & OOM \\
       CF$^2$ & 4.053 & 163.414 & 946.235 & & 7.047 & 263.659 & 1688.145 \\
       CLEAR \textit{(pre-train)} & 0.002 & OOM & OOM & & 0.018 & OOM & OOM\\
       UNR-Explainer & 0.781 & 28.725 & 213.481 & & 1.801 & 39.719 & 476.294 \\
       C2Explainer & 3.552 & 166.961 & OOM & & 5.170 & 185.426 & OOM \\
       \hline
       
    \end{tabular}
    }
    
    \label{tab:comp2}
\end{table}

\section{Conclusion}

This work presents a comprehensive and reproducible benchmarking study of counterfactual explainers for recommender systems. We introduce a unified evaluation protocol that encompasses different explanation formats, evaluation levels, and perturbation scopes, thereby mitigating fragmentation of the evaluation process in prior studies and enabling systematic comparison. Notably, we extended the evaluation paradigm beyond the conventional item-level setting to incorporate, for the first time, a rigorous list-level analysis. Through extensive experiments on multiple datasets and recommender architectures, we provided a holistic evaluation of explanation effectiveness, sparsity, robustness, and computational complexity. The results reveal a consistent trade-off between effectiveness and sparsity under the explicit format, as well as varying degrees of stability when explaining items at different positions in ranked recommendation lists. Moreover, restricting perturbations to $k$-hop subgraphs yields meaningful explanations while limiting redundant interactions from the full user-item graph. Based on these findings, we recommend that future work should: (i) evaluate performance on both item level and list level, (ii) report sparsity and runtime alongside effectiveness, and (iii) explicitly specify the perturbation scope for graph-based explainers.

\pagebreak

\bibliographystyle{ACM-Reference-Format}
\balance
\bibliography{refs}

\end{document}